# EVOLUTION OF MASS OUTFLOW IN PROTOSTARS

Dan M. Watson[1], Nuria P. Calvet[2], William J. Fischer[3,4], W.J. Forrest[1], P. Manoj[5], S. Thomas Megeath[3], Gary J. Melnick[6], Joan Najita[7], David A. Neufeld[8], Patrick D. Sheehan[9], Amelia M. Stutz[10] & John J. Tobin[11].

## ABSTRACT

We have surveyed 84 Class 0, Class I, and flat-spectrum protostars in mid-infrared [Si II], [Fe II] and [S I] line emission, and 11 of these in far-infrared [O I] emission. We use the results to derive their mass outflow rates, $\dot{M}_w$. Thereby we observe a strong correlation of $\dot{M}_w$ with bolometric luminosity, and with the inferred mass accretion rates of the central objects, $\dot{M}_a$, which continues through the Class 0 range the trend observed in Class II young stellar objects. Along this trend from large to small mass-flow rates, the different classes of young stellar objects lie in the sequence Class 0 – Class I/flat-spectrum – Class II, indicating that the trend is an evolutionary sequence in which $\dot{M}_a$ and $\dot{M}_w$ decrease together with increasing age, while maintaining rough proportionality.

The survey results include two which are key tests of magnetocentrifugal outflow-acceleration mechanisms: the distribution of the outflow/accretion branching ratio $b = \dot{M}_w / \dot{M}_a$, and limits on the distribution of outflow speeds. Neither rule out any of the three leading outflow-acceleration, angular-momentum-ejection mechanisms, but they provide some evidence that disk winds and accretion-powered stellar winds (APSWs) operate in many protostars. An upper edge observed in the branching-ratio distribution is consistent with the upper bound of $b = 0.6$ found in models of APSWs, and a large fraction (31%) of the sample have branching ratio sufficiently small that only disk winds, launched on scales as large as several AU, have been demonstrated to account for them.

---

[1] Department of Physics and Astronomy, University of Rochester, Rochester, NY 14627-0171. For email correspondence: dmw@pas.rochester.edu.
[2] Department of Astronomy, University of Michigan.
[3] Department of Physics and Astronomy, University of Toledo.
[4] NASA Goddard Space Flight Center, Greenbelt, MD
[5] Tata Institute of Fundamental Research, Mumbai, India.
[6] Harvard-Smithsonian Center for Astrophysics, Cambridge, MA
[7] National Optical Astronomy Observatory, Tucson, AZ.
[8] Department of Physics and Astronomy, Johns Hopkins University, Baltimore, MD.
[9] Steward Observatory, University of Arizona, Tucson, AZ.
[10] Max Planck Institute for Astronomy, Heidelberg, Germany.
[11] Leiden Observatory, The Netherlands.

## 1. INTRODUCTION

Outflows and accretion in young stellar objects (YSOs) have long been thought to be linked to one another, and both to involve mass flow along poloidal magnetic fields which thread the central object and its surrounding disk (Hartmann & MacGregor 1982, Ghosh & Lamb 1978). Models which describe the flows best are magnetocentrifugal, in which excess angular momentum is carried away by the outflow to permit the rest of the disk-accretion flow to be channeled to the central object. The three classes of these models which have received the most attention are X winds (Shu et al. 1988, 1994, 2000; Najita & Shu 1994; Shang, Li & Hirano 2007; Cai et al. 2008), disk winds (Pelletier & Pudritz 1992 [henceforth PP92]; Wardle & Königl 1993; Königl & Pudritz 2000; Pudritz et al. 2007; Königl, Salmeron & Wardle 2010; Salmeron, Königl & Wardle 2011) and accretion-powered stellar winds (APSW; Matt & Pudritz 2005, 2008a, 2008b [henceforth MP08]; Romanova et al. 2005). The differences between these mechanisms are several, but one important distinguishing feature is the location of footpoints: the distance from the rotational axis at which the open magnetic field lines, which accelerate the bulk of the outflow, are frozen. In APSW models the footpoints are on the central object, and in disk wind models they are well out in the disk, typically at 0.1-1 AU. The footpoints of X wind models lie in between, at the X point: the radius at which the magnetosphere of the central object truncates the gaseous disk, usually a few Solar radii away from the system center. Angular momentum is extracted from the accretion flow at the footpoints. Even in the nearest star-formation regions, the angular scale of nearly all of these locations is beyond the spatial resolution of current observations (cf. Ray et al. 2007), and direct probes of the magnetic field itself are lacking, so the models must be tested less directly than simply by finding the footpoints in images.

A common feature of the three outflow mechanisms is the prediction that mass outflow rate $\dot{M}_w$ scales with the mass accretion rate $\dot{M}_a$ of the central object: $\dot{M}_w = b\dot{M}_a$. The precise value of the branching ratio $b$ depends upon the leverage exerted by the magnetic fields,[12] which in turn is related to the radial position of the footpoints at which the lines of ***B*** are frozen in the disk. For a given accretion rate, and the associated rate at which angular momentum needs to be carried off by the outflow, the mass outflow rate is inversely related to the footpoint radius. Each class of model can be consistent with a range of values of *b*. These ranges overlap, but larger values of *b* are usually required from APSWs and X winds, while disk winds, with a wider range of lever-arms available, can accommodate smaller values. It is, of course, possible that all three mechanisms are always at work in all systems, since in general there are magnetic field lines anchored in all three domains; the question is which one dominates the removal of the last of the angular momentum from the accretion flow, and leads to T Tau stars rotating as slowly as observed.

---

[12] The lever arm in a magnetocentrifugal torque is effectively the so-called Alfvén radius, defined to be the radius at which the poloidal component of the outflow velocity is equal to the Alfvén wave speed; this is generally larger than the radius of the associated footpoint.

In turn, the branching ratio *b* is integral to protostellar evolution. As it describes the branching of mass from the envelope, through the disk, to the central star on one hand and the outflow on the other, its value is important in the understanding of the efficiency of collapse and the magnitude of feedback, to the envelope and to the surroundings.

Observational characterization of *b* has focused on kinematics and energetics of the outflowing and accreting gas. In Class II YSOs – those YSOs which are inferred to lack envelopes – mass outflow rates can be measured *via* certain visible and infrared forbidden-line luminosities. Frequently used are [O I] 630 nm (Cabrit et al. 1990; Edwards et al. 1993; Hartigan, Edwards & Ghandour 1995, henceforth HEG95) and [O I] 63.2 $\mu$m (Werner et al. 1984; Cohen et al. 1988). These lines are bright in dissociative, nonmagnetic – *J* type (Draine 1980) – shocks driven in the outflowing material, as it decelerates upon its first encounter with ambient material (Hollenbach & McKee 1979; 1989, henceforth HM89). Such "wind" shocks often lie rather close to the central object, so observations of longer-wavelength lines, like [O I] 63.184 $\mu$m or the mid-infrared lines of [Si II] or [Fe II], may help overcome significant extinction in Class 0, Class I and flat-spectrum YSOs. Mass outflow rates can be derived from the luminosities of these lines because they are the dominant coolants, or good proxies for the dominant coolant, over well-defined temperature range in the postshock gas (Hollenbach 1985). For example, each particle crossing the $T = 5000$ K isotherm in a *J* shock carries energy of order $5000 \, \text{K} \times k_B$, and almost all of this energy is radiated as collisionally-excited [O I] 63.2 $\mu$m line radiation as the post-shock column cools below this temperature; thus the line luminosity is proportional to the rate at which electrons, or atoms, pass through the shock (Hollenbach 1985; Werner et al. 1984; HM89). Except in objects with very large X-ray luminosity, both the high- and low-velocity components (HVCs and LVCs) of line emission may originate in the outflow's *J* shocks (Hollenbach & Gorti 2009), so one may estimate mass outflow rates from observations in which the spectral-line velocities are unresolved. This is the most direct, precise and accurate method of deriving YSO mass outflow rates; it suffers little from extinction if mid- and far-infrared lines are used, and is free from many uncertainties which afflict the derivation of protostellar mass-loss rates with millimeter/submillimeter molecular lines, such as those of entrainment fraction, opacity, subthermal excitation and molecular relative abundance.

Accretion rates for the T Tau stars in Class II YSOs are usually derived either from the luminosity of hydrogen recombination emission lines – in much the same way as outflow rates from [O I] – or from the degree of spectral-line veiling, or from the degree of ultraviolet excess (e.g. HEG95; Gullbring et al 1998, henceforth G+98; Muzerolle, Hartmann & Calvet 1998, henceforth MHC98). In objects with substantial envelopes, extinction makes these methods extremely difficult. By the same token, however, such objects are in early stages of evolution, in which system luminosity is dominated by accretion power. Thus their mass accretion rates can be estimated from their bolometric luminosities (e.g. Evans et al. 2009, henceforth E+09), which are much more easily obtained.

In the Class II YSO domain, the largest collection of low-mass objects in which visible spectral lines and veiling have been used to determine $b$ is that of HEG95 and G+98. These authors have shown that the expected linear correlation of $\dot{M}_w$ and $\dot{M}_a$ is obtained for $\dot{M}_a = 10^{-7} - 10^{-10} M_\odot$ year$^{-1}$. Similar results obtained for Herbig Ae/Be systems by Corcoran & Ray (1998) indicate that magnetocentrifugal outflow acceleration probably also works in more massive systems.

The Spitzer Space Telescope (Werner et al. 2004) put the low-mass protostellar counterparts of these objects within reach of these sorts of energy-based determinations of the mass flow rates. In this article we present mid-infrared spectroscopic observations of [Si II], [Fe II] and [S I] emission in a large sample of protostars in nearby YSO associations. By comparison to our previous [O I] 63.2 $\mu$m observations we will show that these lines probe the same regions within post-*J*-shock gas as the dominant coolant, [O I], and serve as excellent proxies for [O I]. Together with bolometric luminosities determined for these objects from their 2MASS, *Spitzer*, *Herschel* and ground-based submillimeter broadband spectral energy distributions, the results enable examination of the $\dot{M}_w - \dot{M}_a$ correlation, extending through the protostellar domain.

## 2. OBSERVATIONS

We selected 84 protostars, lying within nine nearby molecular cloud complexes, from the *Spitzer* broadband mid-infrared surveys by the c2d Legacy team (E+09) and the IRAC-IRS Young Clusters team (Megeath et al. 2004). First we demanded that targets have near- to mid-infrared spectral index,

$$\alpha = \log(\lambda_2 F_{\lambda 2}/\lambda_1 F_{\lambda 1})/\log(\lambda_2/\lambda_1) \quad , \tag{1}$$

characteristic of Class 0, I or flat-spectrum objects; that is, $\alpha \geq -0.3$, with $\lambda_2 = 24 \; \mu m$, $\lambda_1 = 2.2, 3.6$ or $4.5 \; \mu m$, and $F_\lambda$ is the continuum flux per unit wavelength interval. Next we demanded that the objects appear compact in *Spitzer* and 2MASS images, to avoid the necessity of extensive mapping. We expected that this restriction would bias the sample toward face-on orientation and/or dynamically-young outflows, but would not in itself yield a sample unrepresentative of protostars.[13] Many of the targets have bipolar outflows, observed in millimeter-wavelength molecular lines, though none are seen to have long, collimated jets. Third, we demanded that the targets be brighter than 7 mJy in the continuum at $\lambda = 24$ $\mu$m; this corresponds roughly to a luminosity of approximately $0.01 L_\odot$ for Class I YSOs in the nearest star-formation regions. Finally we ensured that the 24-$\mu$m luminosity function of the targets was sampled with fair uniformity. In Table 1 we list the sample objects, together with the key parameters that can be derived from their broadband spectral energy distributions: spectral index $\alpha$, bolometric luminosity and temperatures $L_{bol}$ and $T_{bol}$ as defined by Myers & Ladd (1993), and YSO class as defined by Wilking & Lada (1983), André, Ward-Thompson & Barsony (1993), and Greene et al. (1994).

---

[13] It is difficult to determine protostellar axis orientation precisely. For the present targets which lie in the Orion clouds, SED modeling (Furlan et al. 2015, henceforth F+15) yields a median axis orientation of *i* = 63°; not far from that expected in a randomly oriented population (60°).

We used the *Spitzer* Infrared Spectrograph (IRS; Houck et al. 2004) for these observations, which comprised *Spitzer* Programs 50560 and 50633, carried out in 2008-2009. Each target was observed with the IRS LH spectrograph ($18.7-37.2$ $\mu$m, $\lambda/\Delta\lambda \cong 600$). Though long enough to verify one dimension of the expected compact sizes of our targets, the LH spectrograph slit is too short to allow accurate subtraction of sky emission if only on-target observations are available. We used IRS Fixed Cluster Mode to cover the target and also two "sky" positions, symmetrically placed on opposite sides of the target, with angular separation (30-120 arcsec) and orientation chosen when necessary to avoid bright nebulosity apparent on 2MASS and *Spitzer*-IRAC images. Each observation, target or sky, consisted of four pairs of 60-second integrations, with the pointing center nodded between two positions $\pm 1/3$ of the slit length from the slit's center.

We extracted our spectra from the *Spitzer* Science Center S18.7 pipeline basic calibrated data using the SMART software package (Higdon et al. 2004). Permanently bad and "rogue" array pixels were corrected by interpolation of signals from normally-functioning pixels, adjacent to the bad ones in the spectral-dispersion direction, before signal from the full $2\times 5$-pixel slit was extracted. The average spectrum of the two sky positions was subtracted from the target spectrum to remove this emission to first order. In an identical fashion, we also extracted LH spectra of two photometric standards: the star $\xi$ Dra (K2III), and the active galaxy nucleus Markarian 231, which we use in the spectral ranges $18.7-30$ and $30-37.2$ $\mu$m respectively. We divided each spectrum's extracted orders at both nod positions by the extracted spectrum of the appropriate standard. Each quotient was then multiplied by the template spectrum of the photometric standard that was used in the division. The resulting nods were averaged to obtain the final spectra. Uncertainties were calculated from the standard deviation of the mean among the individual scans. In each case the resulting uncertainty is similar to that propagated from the measured instrumental sensitivity, and to half the difference between corresponding nod spectra, as expected. In Figure 1 we show the resulting spectra for three typical objects in our sample. Many molecular and atomic spectral lines are typically detected, with several fine-structure lines often prominent: [Si II] 34.815 $\mu$m; [Fe II] 24.519, 25.988, and 35.349 $\mu$m; and [S I] 25.249 $\mu$m.

We extracted [Si II], [Fe II] and [S I] line fluxes from the final spectra by $\chi^2$-minimization fits of the instrumental spectral profile and second-order polynomial continuum in 15-20 channel (pixel width in the spectral direction) ranges centered on the spectral lines. Uncertainties in each channel of the final spectra were propagated to the uncertainties of the line fluxes *via* the covariance matrix elements associated with each fit. The resulting measurements are presented in Table 3. We compare these fluxes with the instrumental sensitivity and flat-field limits in Figure 2. For the strongest line of each of these three species, [Si II] 34.8 $\mu$m, [Fe II] 26.0 $\mu$m, and [S I] 25.2 $\mu$m, the line-to-continuum ratio is well above the flat-field limit in the vast majority of cases, so there is no tendency in these lines for the flat-field limit artificially to produce a positive correlation between line flux and continuum flux or luminosity, nor an upper or lower edge to the flux distribution.

In our observations, the [Si II], [Fe II] and [S I] emission appears always to be both spatially and spectrally unresolved. Emission comes either exclusively or dominantly from an unresolved source coincident with the protostar; that is, within an $11"\times 23"$ box centered on the target. The linewidths are all much smaller than the spectral resolution, $c\Delta\lambda/\lambda$ = 500 km sec$^{-1}$ (FWHM).

We have previously reported [O I] 63.184 $\mu$m *Herschel*-PACS observations of eleven objects on the present list (Manoj et al. 2013). For reference these observations are summarized in Table 2. The spatial resolution of *Herschel* at this wavelength (10 arcsec) is quite similar to the *Spitzer* IRS LH slit width (11 arcsec) so most of our targets are also expected, and observed, to be spatially unresolved in the [O I] 63.2 $\mu$m observations. In each case, the line profile appears to be slightly but significantly broader than the instrumental profile, which is 102 km sec$^{-1}$ wide (FWHM). A rough estimate of the intrinsic velocity dispersion can be obtained by presuming the intrinsic and instrumental profiles both to be Gaussian, and subtracting in quadrature. This produces a median intrinsic velocity width (FWHM) of 60.6 km sec$^{-1}$. The range of linewidths we infer is similar to that observed for the low-velocity components of Class II YSO outflows (e.g. HEG95).

We detect mid- and far-infrared continuum emission in each of our targets, which arises near their protostellar cores, along lines of sight through their cold envelopes. If within the envelopes there are substantial concentrations of cold, neutral S or O, or ionized Si or Fe, absorption of the continuum emission by these atoms and ions could artificially reduce the line flux we detect when integrating over velocity ranges of absorption and emission from both hot and cold gas near the protostar. There is no hint of such self-absorption in our [O I] spectra, so any actually present would require much higher spectral resolution to detect. In our experience, such self-absorption has only been observed, at spectral resolution similar to ours, in two particularly high-mass molecular cores, DR21 and NGC 6334V (Poglitsch et al. 1996, Kraemer et al. 1998). No [O I] self-absorption is detected at very high spectral resolution in the massive protostar Orion-IRc2 (Boreiko & Betz 1996). Thus we lack a direct estimate its magnitude in our low-mass protostar sample. However, the atomic column densities in the envelopes of our sample are several orders of magnitude smaller than those toward Orion-IRc2, DR21 or NGC 6334V, judging by the typical total extinction we infer toward our targets (§3.2), so the related [O I] absorption should be insignificant. The absorption would span only a fraction of the velocity spread of matter within the envelope, usually about 1-2 km; a very small fraction of the line width, if our [O I] observations are used to judge. We therefore neglect this effect in first approximation, but note that we might thereby underestimate [O I] line fluxes, and thus the outflow rates, slightly in some objects. As sulfur, iron and silicon are much less abundant than oxygen, this effect would be even smaller in our [S I], [Fe II] and [Si II] observations.

## 3. ANALYSIS

*3.1. J-type wind shock origin of fine-structure line emission in protostars.*

Our observations of [Fe II], [Si II], [S I] and [O I] lines are described well by models of the *J*-type shocks expected in the outflows from protostars. The emitting medium fits the description of post-*J*-shock gas in other ways as well, such as its relatively high density independent of shock models, and its elevated $Fe^+/O$ and $Si^+/O$ abundance ratios.

In Figure 3 we plot observed [Si II], [Fe II] and [S I] line fluxes against the [Fe II] 26.0 $\mu$m flux. Four of the five lines – all but [Fe II] 24.5 $\mu$m – are covered in the *J* shock models by HM89. The distribution of the intensity ratios of these lines is in good agreement with the HM89 results, for preshock hydrogen-atom densities in the range $n_0 = 10^3 - 10^5$ cm$^{-3}$ and shock speeds of $v_S = 30 - 150$ km sec$^{-1}$. Because the peak temperatures reached in *J*-type shocks from YSO outflows lie in the $T = 10^4 - 10^5$ K range, and have cooled before [O I], [Si II], [Fe II] and [S I] lines begin emitting strongly at about $T = 5000$ K, none of the line ratios are exquisitely sensitive to shock speed within the $30 - 150$ km sec$^{-1}$ range. A shock speed of $v_S = 40$ km sec$^{-1}$ divides the [Si II] 34.8 $\mu$m/[Fe II] 26.0 $\mu$m and [S I] 25.2 $\mu$m/[Fe II] 26.0 $\mu$m detections roughly in half. We will take this median speed to be typical of the sample, noting that the velocity is similar to the estimated intrinsic half-widths of the [O I] line profiles (Table 2). Detailed modelling would yield precise determination of preshock density, and useful determination of shock speed, for the targets in which we detected a majority of the lines, which we leave to a future article.

A crude but model-independent gauge of the post-shock conditions can be obtained from a single-component analysis of the [Fe II] intensity ratios, which we present in Figure 4. The [Fe II] 35.3/26.0 $\mu$m and 24.5/26.0 $\mu$m line ratios depend differently on the density and temperature of the single component, and converge to within 8% of the regression values at densities $n_e \gtrsim 10^5$ cm$^{-3}$ and temperature $T \approx 1600$ K. These values are well within the range expected in gas compressed by a *J* shock, and are quite different from those ordinarily found in photodissociation regions (PDRs) associated with molecular clouds; here is strong support for a shock origin of the line emission we detect.

The close agreement of the single-component line flux ratios to the regression values also encourages us in the neglect of extinction, which we have done tacitly above when applying the unextinguished HM models and noting their good fit to the observations. If the 8% difference in the curves of Figure 4 is taken as an upper bound to reddening between wavelengths 35.3 and 24.5 µm, the extinction relations of McClure (2009) imply typical extinction of $A_V = 6-9$ toward the emitting regions of our targets, and thus a reddening between the [O I] 63.2 $\mu$m and [Fe II] 26.0 $\mu$m lines of 0.13-0.15 mag at the most (13-15%), similar to or smaller than the experimental uncertainties. It would appear that the emission regions in our targets are generally extinguished negligibly at the wavelengths of the [Fe II], [Si II] and [O I] lines.

The flux ratios we observe for [Si II] 34.8 $\mu$m/[O I] 63.2 $\mu$m and [Fe II] 26.0 $\mu$m/[O I] 63.2 $\mu$m appear to be essentially constant, as we show in Figure 5. This is expected, according to HM89: over the preshock density and shock speed ranges $10^3 - 10^5$ cm$^{-3}$ and $30-150$ km sec$^{-1}$, the model intensity ratios [Si II] 34.8 $\mu$m/[O I] 63.2 $\mu$m and [Fe II] 26.0 $\mu$m/[O I] 63.2 $\mu$m are essentially constant (see Figure 5), independent of preshock density and shock speed. The reasons for the constancy of line ratios are that, for temperatures below 5000 K in the post-shock gas, O, Si, and Fe stay predominantly in the form O, Si$^+$, and Fe$^+$; that all three lines, [O I] 63.2 $\mu$m, [Si II] 34.8 $\mu$m, and [Fe II] 26.0 $\mu$m, are ground-state transitions with similar critical densities for collisional de-excitation; and that the temperature in most of the post-shock column is much greater than the upper-state $E/k_B$ values: the line emission coefficients behave the same way as the postshock gas cools and compresses.

We also see in Figure 5 that the fit of *J*-type shock models to the [Fe II], [Si II] and [O I] data yields estimates for the gas-phase relative iron, silicon and oxygen abundances: $Fe^+/O = 1.9 \times 10^{-3}$ and $Si^+/O = 6.7 \times 10^{-3}$. As oxygen is depleted very little in the interstellar medium (e.g. Jensen, Rachford & Snow 2005), and as iron and silicon are almost entirely in singly-ionized form in the bright parts of the shocks, these abundance ratios indicate that at least ~20% and ~3.5% respectively of the solar-system abundances of Si and Fe (Lodders & Palme 2009) are in the gas phase in the emitting regions of our targets. This would be on the high end of gas-phase relative abundances in cold interstellar-medium sightlines such as that toward ζ Oph, for which 5% and 1% respectively are typical (e.g. Miller et al. 2007). That there may be a factor of four more than the usual gas-phase concentrations of Si and Fe in the emitting regions of our targets supports the idea that these regions are post-shock gas, in which a moderate amount of grain sputtering has taken place.

Thus we can take the emission to arise in the shocks driven by these protostars' outflows in relatively dense material near their base, as is generally the case with the visible and near-infrared forbidden-line emission from the base of the outflows of Class II objects (e.g.HEG95).

*3.2. Negligible fine-structure line emission from photodissociation regions or embedded disks protostars.*

Besides shocks, at least two otherwise plausible, alternative origins of the line emission need to be considered: relatively low density PDRs associated with the protostellar envelopes or their surroundings; and the higher-density photodissociated gas on the surfaces of our targets's embedded disks. Our information on spatial and velocity structure of the emission region cannot rule out these origins decisively, but the observed fine-structure line fluxes, and their ratios, can.

Regions containing YSOs frequently exhibit bright [Si II] and [O I] emission from PDRs (e.g. Haas, Hollenbach & Erickson 1986, Liseau, Justtanont & Tielens 2006). But [Fe II] emission from PDRs is generally very weak by comparison. According to the models by Kaufman, Wolfire & Hollenbach (2006), [Si II] 34.8 $\mu$m/[Fe II] 26.0 $\mu$m lies in the range 60-200 for PDRs and H II regions. According to the observations of PDRs by Okada et al. (2008), [Si II] 34.8 $\mu$m/[Fe II] 26.0 $\mu$m lies in the range 40-200, and these lines are accompanied by [Fe III] 22.93 μm emission which is typically similar in intensity to [Fe II] 26.0 μm. In contrast, the [Si II] and [Fe II] lines are always similar in brightness in our present survey sample. None of our targets exhibit [Si II] 34.8 $\mu$m/[Fe II] 26.0 $\mu$m near the PDR range, nor exhibit extended [Si II] sky emission which is significant compared to the compact sources associated with our targets. We also detect no hint of [Fe III] in any of our targets; typically we obtain [Fe III] 22.9 $\mu$m/[Fe II] 26.0 $\mu$m < 0.05 (3$\sigma$) from our spectra. Our [O I] line observations indicate that, at least for the eleven objects in Table 3, the line widths, 40-90 km sec$^{-1}$ (FWHM), are much larger than observed or expected in PDRs ($\lesssim$ 10 km sec$^{-1}$; e.g. Boreiko & Betz 1997, Ossenkopf et al. 2013). We therefore reject the possibility of significant PDR contribution to our observations.

Ultraviolet starlight excites [O I] emission from the surfaces of disks, as seen in *Herschel* observations of Class II YSOs. But this emission is very faint compared to what we observe. The median [O I] 63.2 $\mu$m luminosity of our targets is larger than that obtained by the *Herschel*-GASPS survey (Aresu et al. 2014) by a factor of 270, though the median bolometric luminosity of our targets is larger than the GASPS targets by only a factor of 7. Even if we neglect the additional, internal ultraviolet extinction to which illumination of an embedded disk may be subject, this comparison suggests that at most an embedded disk would contribute only a few percent of the [O I] luminosity of our targets. And neither [Fe II] nor [Si II] has been detected in any of the ~2000 Class II YSOs observed with *Spitzer*-IRS. We therefore rule out the embedded disks as significant contributors to the line emission we observe in this survey.

*3.3. Derivation of protostellar mass outflow rate from [Fe II], [Si II] or [O I] luminosity.*

The total mass outflow rate through *J*-type shocks can be obtained from the cooling luminosity, as described above (§1); in turn the cooling luminosity is given accurately by the luminosity of [O I] 63.2 μm (Hollenbach 1985; HM89). In Figure 5 we see that [Si II] 34.8 μm and [Fe II] 26.0 μm both serve as accurate proxies for [O I] 63.2 μm. Scaling from the HM89 [O I]–$\dot{M}_w$ relation, we can compute the total cooling luminosity, and obtain the mass flow rate through the *J* shocks, from the [Si II] and [Fe II] line fluxes as well as from [O I]:

$$\begin{aligned}
\dot{M}_w &= 8.1 \times 10^{-5} L\big([\text{O I}]\ 63.2\ \mu\text{m}\big) \frac{M_\odot\ \text{year}^{-1}}{L_\odot} \\
&= 9.8 \times 10^{-4} L\big([\text{Si II}]\ 34.8\ \mu\text{m}\big) \frac{M_\odot\ \text{year}^{-1}}{L_\odot} \\
&= 1.4 \times 10^{-3} L\big([\text{Fe II}]\ 26.0\ \mu\text{m}\big) \frac{M_\odot\ \text{year}^{-1}}{L_\odot}
\end{aligned} \tag{2}$$

Formally these are upper limits to the mass loss rate, as internal shocks from reflection within the jet can lead to fine-structure line emission in addition to that produced by the first deceleration of the outflow. In practice, this is a small effect (Cohen et al. 1988, HEG95) and we will neglect it henceforth. In the numerous cases in which both the [Si II] and [Fe II] lines are detected, we averaged the resulting values of $\dot{M}_w$, and added their uncertainties in quadrature.

We plot each survey object's resulting mass-outflow rate $\dot{M}_w$ against its bolometric luminosity $L_{\text{bol}}$ (from Table 1) in Figure 6. For comparison we also plot $\dot{M}_w$ and $L_{\text{bol}}$ for selected Class II objects, for which HEG95 determined $\dot{M}_w$ from [O I] 630 nm observations. We obtain thereby a distribution with three notable properties. First, mass-outflow rate for the protostars is strongly and linearly correlated with bolometric luminosity. This trend among the protostars is not followed by the Class II objects. Second, the Class 0 objects tend to have larger $\dot{M}_w$ and $L_{\text{bol}}$ than Class I and flat-spectrum objects; that is, the expected evolution of outflow rate with envelope predominance is evident. The principal components of the data for the separate YSO classes indicate that Class I and flat-spectrum objects are not distinguished from one another by range of $L_{\text{bol}}$ and $\dot{M}_w$, but that Class 0 is distinctively different from the other protostars. The first principal components also indicate that the protostar population exhibits its greatest variation in the same direction as the trend; that is, in the "evolution" direction. Third, the distribution of protostars seems to have a sharp upper edge in $\dot{M}_w$ for given $L_{\text{bol}}$. The lower bound, dominated by upper limits to $\dot{M}_w$, is less abrupt.

*3.4. Mass outflow and mass accretion rates in protostars*

Disk-star accretion rates $\dot{M}_a$ have been measured for the Class II objects plotted in Figure 6 by G+98. What of the protostars in our sample? As described in §1, it is common to assume that accretion power dominates system luminosity for young stellar objects with envelopes, and to estimate $\dot{M}_a$ phenomenologically from $L_{\text{bol}}$ *via* a relation such as that used by Enoch et al. (2009):

$$\dot{M}_a = \frac{2 L_{\text{bol}} R_p}{G M_p} \cong \frac{20 L_{\text{bol}} R_\odot}{G M_\odot} \quad , \tag{3}$$

where $R_p$ and $M_p$ are the radius and mass of the central star. This scaling of $L_{\text{bol}}$ to $\dot{M}_a$ gives a good account of Class 0/I evolution for the nearby clouds in the c2d survey (Enoch et al. 2009, E+09), even though it includes two major simplifying assumptions: that all protostars have the same radius-to-mass ratio, and that the observed $L_{\text{bol}}$ is approximately the same as the total luminosity $L$ emitted by the object in all directions. Better estimates for $R_p/M_p$ and $L$ must be obtained from detailed models, for which we would lack sufficient constraints for much of our survey.

In the present case the observation of a linear correlation between $\dot{M}_w$ and $L_{\text{bol}}$ for the protostars in Figure 6, and the expectation from magnetocentrifugal outflow models of a linear relationship between $\dot{M}_w$ and $\dot{M}_a$, makes plausible the use of Equation (3) to obtain accretion rates for our sample. The luminosities of the youngest of our targets, the Class 0 objects, are very probably dominated by accretion power. Many Class I objects should also be accretion-powered. Some typical Class I protostars have be shown – *via* observations of infrared hydrogen recombination lines and veiling of photospheric lines, and a difficult correction for envelope extinction – to be powered mostly by accretion (Prato et al. 2009; see also Nisini et al. 2005), in contrast with some earlier findings (e.g. MHC98) that the accretion luminosities of many Class I objects are much less than their total luminosities[14]. The chances increase, as a protostar passes through the

---

[14] Part of the contrast is due to more comprehensive envelope-extinction correction in recent work, and part from early regard of some deeply-embedded objects as Class I and envelope-dominated, which today are classified as Class II and disk-dominated – and for accretion luminosities are expected to be small fractions of $L_{\text{bol}}$. For example, nine of the ten Class I objects in Ophiuchus for which MHC98 measured accretion luminosities were observed by *Spitzer*-IRS, with the result that seven were classified as disk-dominated, and two as envelope-dominated, according to spectral indices which depend very weakly on extinction (McClure et al. 2010). Of the two objects which are envelope-dominated by these

developmental stages represented by Class I, for the central star's intrinsic luminosity to exceed the accretion luminosity. By the stage represented by most flat-spectrum objects, power from these other sources probably exceeds that from accretion (E+09, F+15). So we shall use Equation (3) to give estimates for $\dot{M}_a$ to good approximation for Class 0 objects. In the absence of deep infrared spectroscopic observations of our current sample from which accurate accretion rates can be obtained, we use Equation (3) with caution for Class I objects. We take results of Equation (3) as upper limits to $\dot{M}_a$ for flat-spectrum objects.

In Figure 7 we plot the accretion and outflow rates, $\dot{M}_a$ and $\dot{M}_w$, for our protostar sample and for the Class II YSOs with $\dot{M}_a$ from G+98 and $\dot{M}_w$ from HEG95. In contrast to Figure 6, the Class II objects and the protostars now appear to follow a common trend, in which the protostars comprise a smooth extension of the correlation reported by HEG95 for Class II objects. Also apparent in Figure 7 is the linear upper boundary to the distribution of points, inherited from Figure 6: its position is consistent with a branching ratio $b = \dot{M}_w/\dot{M}_a = 0.6$, the value obtained by MP08 as the maximum of $b$ for accretion-powered stellar winds. Linear regression to the protostellar data yields a typical branching ratio $b = 0.089$, similar to the canonical, theoretical, value of 0.1 (e.g. PP92). The second principal components, which would be more sensitive to the spread of the branching ratio $b$ and therefore the footpoint radius, do not appear to vary by much among the four YSO populations. Clearly, then, the evolutionary sequence Class 0 – Class I/flat-spectrum – Class II is evident in the trend: mass outflow and accretion decrease in roughly the same proportion as the population evolves.

The scatter along the trend is substantial and real; it is interesting to wonder whether it can be linked to time variability. All young stellar objects are variable on a wide range of time scales, in ways that are often suspected to originate in variation in the disk-star accretion flow. None of our targets are currently FUors or EXors; none have exhibited large variability in infrared luminosity during the *Spitzer* cryogenic mission (e.g. Megeath et al. 2012, F+15). However, variability provides an additional, unknown uncertainty. The comparison of data and HM89 models in Figure 3, and the intrinsic [O I] line widths estimated in Table 2, suggest that the shock speed is typically $v_s \sim 40$ km sec$^{-1}$. Let us conservatively take the outflow speed $v_w$ to be the same as the shock speed. Generally, our targets are spatially unresolved by the 11 arcsec beam of *Spitzer*-IRS LH and the 10 arcsec beam of *Herschel*-PACS, corresponding to a maximum extent of about $R = 770$-$2300$ AU from the central object. Thus the outflowing material was ejected from the vicinity of the central object $\Delta t = R/v_w = $ 100-300 years, at the most, before it emitted the light we detect. The time scale for the response of bolometric luminosity or H recombination-line emission to a change in accretion rate can be much shorter than this. So there is additional

---

criteria, one, WL 17, is in fact heavily accretion dominated and the other, WL 12, is not, according to MHC98.

uncertainty in our values of *b* due to flow-rate variations on time scales less than a few hundred years. It is highly improbable that any of our targets was involved in an untimely FU Ori event, for which the recurrence time scale, $10^3 - 10^4$ years, is very long compared to $\Delta t$. Similarly the rarity of EXors and V1647 Ori-like outbursts, and the possibility that protostars have not yet developed the instability thought to produce such outbursts (Audard et al. 2014), makes it unlikely that this effect has occurred in our sample, despite a recurrence time considerably less than $\Delta t$. But even non-outbursting YSOs often exhibit stochastically-variable accretion rates with amplitudes which are appreciable fractions of the time-average rates (e.g. Eisner et al. 2015).

## 4. DISCUSSION

*4.1. Observed outflow-accretion branching ratios, and those in disk wind, X wind, and APSW models*

For greater ease of comparison to the magnetocentrifugal-acceleration outflow models, we project the data in Figure 6 and Figure 7 into an empirical distribution function (EDF)[15] of the branching ratio $b = \dot{M}_w / \dot{M}_a$, in Figure 8. The resulting EDF increases steadily from small values of *b* up to *b* ~ 0.5-0.7, and then flattens sharply. Over much of its range, the EDF is very similar to the cumulative distribution function for a lognormal probability distribution, but turns over more sharply at the upper edge, and flattens more gently at lower values. That there is a sharp upper edge, and that its value is close to limiting value *b* = 0.6 in APSWs (MP08), may be evidence that the largest values of *b* within the sample are associated with APSWs.

The observed branching-ratio ranges map into the footpoint ranges involved in the three outflow mechanisms, but not sharply. X-wind models work well in the range *b* = 0.1-1 (e.g. Najita & Shu 1994); off-the-shelf versions of these models would suffice for 50% of the sample, but do not naturally lead to a well-defined maximum in the branching ratio. Published APSW models can reproduce *b* values from about *b* = 0.06 up to the cut-off at *b* ~ 0.6 – encompassing about 56% of the sample – and they account for this cut-off. For about 43% of the sample which lies near the center of the EDF, any of the three mechanisms can consistently explain the observed values of *b*; it is generally this range in which outflow models have been optimized.

But a substantial fraction of the sample has branching ratios well below this range. A good example is HOPS 87 (a.k.a. OMC-3 MMS6), the most luminous object in our study. HOPS 87 is a Class 0 object for which we infer $\dot{M}_a = 2.3 \times 10^{-5} M_\odot$ year$^{-1}$. It possesses an especially compact bipolar outflow seen in submillimeter molecular lines (Takahashi &

---

[15] Since a significant fraction of the *b* results are upper or lower limits, a histogram would suffer from more than the usual ambiguities, so we opt for an EDF, employing Efron's implementation of the Kaplan-Meier estimate (e.g. Feigelson & Babu 2012, p. 268) so as to include the censored data and the information they convey.

Ho 2011), and emits a rich mid-and far-infrared molecular-line spectrum, dominated by CO and $H_2O$ (Manoj et al. 2013; Sheehan et al., in preparation; Manoj et al., in preparation), which is probably associated with the outflow. Yet it shows no detectable [Si II], [Fe II], or [O I] emission; the resulting $3\sigma$ upper limit to HOPS 87's branching ratio is $b < 0.003$. This is much smaller than has ever been demonstrated in X wind or APSW models, but is within the grasp of numerical (Salmeron, Königl & Wardle 2011) and analytical (PP92, Wardle & Königl 1993) disk-wind models. For example, the PP92 model can be scaled from a Class II object and $b$ = 0.27 to a Class 0 object and $b$ < 0.003, with a few minor and plausible adjustments to the parameters. The most significant adjustment is of the effective radius $r_e$ of the disk-wind region from 1.2 AU to a lower limit of > 5 AU, a size scale that will be accessible to ALMA in the near future. Some 31% of the sample lies in this range in which neither X-winds nor APSWs have been demonstrated to work; only disk-wind models currently suffice. In all, disk winds have been demonstrated to accommodate all branching ratios from about $b = 10^{-5}$ (e.g. Wardle & Königl 1993) to about $b$ = 0.3 (e.g. PP92), which would encompass about 74% of the present sample.

*4.2. Kinematics of disk wind, X wind, and APSW models*

The different magnetocentrifugal outflow-acceleration mechanisms differ in outflow speed as well as footpoint position. Even with the limited spectral resolution of our survey, useful kinematic limits can be placed on the models. For example, the profile of each spectral line with IRS LH we detect is fit precisely with the instrumental resolution, $\Delta v \cong 500 \text{ km sec}^{-1}$ FWHM. Given the essentially Gaussian shape of the instrumental responses, this means $(v_w \cos i)^2 \ll (250 \text{ km sec}^{-1})^2$ for every object we detect. Such a range is consistent at any inclination with the spread of shock speeds in the HM89 models within which the data fit – presuming that the shock and wind speeds are the same – and with the velocity dispersion inferred for eight of our targets in [O I] 63.2 $\mu$m ($40-90 \text{ km sec}^{-1}$ FWHM; see Table 2). As for the outflow models, the restriction is hardest to meet with APSW and X winds, but is not fatal to either. APSWs tend to have maximum wind speeds of $v_w = 100-200 \text{ km sec}^{-1}$ (Matt & Pudritz 2008a), requiring just a little help from orientation to match our observations. X winds have speeds typically in the range $v_w = 100-500 \text{ km sec}^{-1}$ (e.g. Najita & Shu 1994). The lowest speeds, however, correspond to the highest values of *b*; any X-wind-dominated objects in our sample which have $b$ ~ 0.1 may require special orientation in order not to broaden the lines past the instrumental profile of IRS. This does not, of course, rule out the possibility that such winds provide a high-velocity component too faint for us currently to detect.

On the other hand, disk winds satisfy the speed restrictions comfortably. Because they involve footpoints with a substantial range of radii, they always exhibit a wide range of outflow speeds. The outflow would tend to be dominated by the outer radii simply because the disk area sampled increases with radius, so the velocity profile features a low-speed core with high-speed wings, as has been noted in the context of the velocity structure of forbidden lines in Class II objects with their low- and high-velocity components (LVCs and HVCs; e.g. HEG95). Observations like those we report here would of course be dominated by the core, as this would dominate the mass flow. For example, in the fiducial, $b$ = 0.27 Class II model by PP92, the speed characteristic of the outflow is $v_w$ = 41 km sec$^{-1}$, though there are speeds as large as 710 km sec$^{-1}$ in material emerging from the innermost parts of the disk. The characteristic speed of this model is a good match to the typical shock speed inferred above from comparison of the data with the HM89 models, and to our [O I] data.

*4.3. Evolution of protostellar mass outflow, mass accretion and outflow force*

The linear $\dot{M}_a - \dot{M}_w$ evolutionary relation depicted in Figure 7 is reminiscent of many molecular-line (usually CO) surveys of protostellar outflows which have revealed a power-law relationship between $L_{\text{bol}}$ and the outflow force $F_{\text{CO}} = d(M_{\text{CO}} v_{\text{CO}})/dt$ (e.g. Wu et al. 2004). Much of the emitting material in molecular outflows is entrained from the surroundings and pushed by the wind whose deceleration we see in the fine-structure lines. Comparison of $F_{\text{CO}}$ with $\dot{M}_w$ for a given outflow can be used to study the entrainment process. The outflow force is calculated from molecular column densities in velocity-resolved line profiles, using transitions of molecules of well-characterized relative abundance, in which the line wings corresponding to the outflow are thought to be optically thin. Usually the power law is not as steep as the one we find for $\dot{M}_a - \dot{M}_w$, but for small ranges of $L_{\text{bol}}$ the relation is often cited as a linear one: $F_{\text{CO}} = aL_{\text{bol}}/c$, where $a$ = 200-400.

We have yet to compile molecular-line observations of our present sample from which to derive $F_{\text{CO}}$. But there are collections of low-mass YSO observations to which ours can be profitably compared, such as the venerable protostellar-outflow sample of Bontemps et al. (1996; henceforth B+96; see also Hatchell, Fuller & Richer 2007, Dunham et al. 2014). B+96 selected Class 0 and I objects in a range of $L_{\text{bol}}$ similar to that in our sample; they determined the outflow force homogeneously from single-dish CO $J$ = 2-1 maps and line wings. For their Class I objects, B+96 found $F_{\text{CO}} \propto L_{\text{bol}}^{0.9}$, close enough to linear that, to facilitate comparison, we have refit their Class I data to a straight line through the origin. Then we applied Equation (3) and the median branching ratio of the linear $\dot{M}_a - \dot{M}_w$ trend obtained for our sample, and found that

$$F_{\text{CO}} = 2.5 \times 10^{-6} M_\odot \text{ km sec}^{-1} \text{ year}^{-1} (L_{\text{bol}}/L_\odot) = \dot{M}_{\text{CO}} v_{\text{CO}}$$
$$= 3.9 \text{ km sec}^{-1} \left(\dot{M}_a / \left[M_\odot \text{ year}^{-1}\right]\right) = 44 \text{ km sec}^{-1} \left(\dot{M}_w / \left[M_\odot \text{ year}^{-1}\right]\right) \quad . \tag{4}$$

The typical wind outflow speed yielded thereby, $v_w = F_{CO}/\dot{M}_w = 44$ km sec$^{-1}$, compares favorably to the typical J shock speed we infer for our sample (§3), to the outflow speeds typically generated in disk-wind models (§4.2), to the intrinsic half-widths we infer in our [O I] 63.2 μm observations, and to the extent of forbidden-line LVCs of Class II objects. B+96 report a typical velocity extent of $v_{CO} \sim 5$ km sec$^{-1}$ from line center for outflows they observe. By this token, typically $(\dot{M}_{CO} - \dot{M}_w)/\dot{M}_{CO} = 0.89$ of the mass seen in their molecular outflows is matter entrained from the surroundings, either envelope or ambient.

## 5. CONCLUSIONS

a. We observe in protostars a strong linear correlation between the bolometric luminosity $L_{bol}$ and the mass outflow rate $\dot{M}_w$. By relating $L_{bol}$ to the mass accretion rate $\dot{M}_a$ onto the central object, we obtain a linear correlation between $\dot{M}_w$ and $\dot{M}_a$. This correlation extends the trend seen in Class II objects through flat-spectrum, Class I and Class 0. The median value of $b = \dot{M}_w/\dot{M}_a = 0.089$ is close to the canonical 0.1. The distribution of $b$ exhibits an edge consistent with the $b \cong 0.6$ maximum seen in APSW models of YSO outflows. Given that only APSW models produce a $b$ distribution with a well-defined upper bound, observation of this edge can be taken as evidence for APSWs in the protostars with large $b$.

b. The distribution of $\dot{M}_a$ and $\dot{M}_w$ is very similar for Class I and flat-spectrum protostars, but these objects are distinct in the correlation from Class 0 and Class II, with Class 0 occupying the high $\dot{M}_a$-$\dot{M}_w$ end of the trend. Thus $\dot{M}_a$ and $\dot{M}_w$ roughly maintain their proportions through the evolution from Class 0 to Class II.

c. The leading magnetocentrifugal-acceleration models have not been optimized at all possible values of $b$; to take the limits of the acceleration mechanisms to be the same as those of the published models may be an inappropriate bias. But if we do take current model results at face value, and suppose that disk winds, X winds and APSWs exhaust the possibilities for outflow-driving mechanisms in YSOs, the observed distribution of mass outflow and accretion rates suggests that disk winds and APSWs dominate the ejection of angular momentum respectively in the objects comprising the tail to low $b$ values, and in those near the cut-off at $b = 0.6$. If indeed the outflows in objects with the smallest $b$ are driven by disk winds, the launching region may be resolvable by ALMA.

For some 43% of the sample, any of the three mechanisms alone can account for the observed values of $b$ according to present models, and the dominant outflow-driving mechanism must be identified on other grounds. The demonstrated domains of operation of disk winds and APSWs overlap; a parsimonious approach to YSO outflow modelling might succeed with just these two mechanisms.

d.  Several lines of reasoning, and limited kinematic evidence from our observations, suggest that the mass loss of protostars is dominated by flows moving at about $v_w \approx 40$ km sec$^{-1}$, similar to the speeds of the LVCs in Class II outflows; contributions to $\dot{M}_w$ from HVCs are small. This suggestion can be tested in some of the objects in our sample by observations with new high-resolution spectrographs on SOFIA.

We are grateful to Ingrid Koch for her help with the IRS data reduction. This work was supported in part by NASA grant NNX14AF79G.

Table 1: broadband infrared properties of our protostar sample.

| Name | J2000 coordinates, degrees | | $d$, pc | $L_{bol}$, $L_\odot$ | $T_{bol}$, K | $\alpha$ | YSO class | Notes |
|---|---|---|---|---|---|---|---|---|
| SSTc2d J032637.5+301528 | 51.65608 | 30.25785 | 250 | 1.00 | 69 | 1.24 | 0 | 1 |
| L1455 IRS4 | 51.93021 | 30.20801 | 250 | 1.70 | 61 | 1.95 | 0 | 1 |
| SSTc2d J032856.6+310737 | 52.23596 | 31.12692 | 250 | 0.01 | 160 | -0.08 | I | 1 |
| SSTc2d J032909.1+312129 | 52.28779 | 31.35808 | 250 | 0.23 | 500 | -0.09 | flat | 1 |
| NGC 1333 IRAS4B | 52.30000 | 31.21893 | 250 | 0.99 | 55 | > 1 | 0 | 1 |
| SSTc2d J032913.6+311358 | 52.30667 | 31.23284 | 250 | 0.80 | 37 | 3.09 | 0 | 1 |
| SSTc2d J032917.2+312746 | 52.32171 | 31.46284 | 250 | 0.40 | 49 | 1.69 | 0 | 1 |
| SSTc2d J032922.3+311354 | 52.34283 | 31.23180 | 250 | | | 0.21 | flat | 1 |
| SSTc2d J033327.3+310710 | 53.36379 | 31.11950 | 250 | 1.70 | 78 | 2.26 | 0 | 1 |
| SSTc2d J034430.3+321135 | 56.12629 | 32.19311 | 320 | 0.01 | 710 | 0.04 | flat | 1 |
| IRAM 04191+15 | 65.50175 | 15.50593 | 140 | 0.28 | | 0.41 | 0 | 1 |
| HOPS 41 | 83.62267 | -5.59519 | 420 | 1.94 | 82 | 1.55 | I | 2 |
| HOPS 32 | 83.64771 | -5.66642 | 420 | 2.01 | 59 | 0.94 | 0 | 2 |
| HOPS 28 | 83.69704 | -5.69885 | 420 | 0.49 | 46 | 1.34 | 0 | 2 |
| HOPS 38 | 83.76967 | -5.62008 | 420 | 0.25 | 59 | 0.94 | 0 | 2 |
| HOPS 40 | 83.78567 | -5.59978 | 420 | 2.69 | 38 | 1.25 | 0 | 2 |
| HOPS 10 | 83.78750 | -5.97433 | 420 | 3.33 | 46 | 0.79 | 0 | 2 |
| HOPS 371 | 83.79338 | -5.92798 | 420 | 0.57 | 32 | 0.00 | 0 | 2 |
| HOPS 91 | 83.82879 | -5.01413 | 420 | 4.15 | 42 | 1.89 | 0 | 2 |
| HOPS 56 | 83.83113 | -5.25906 | 420 | 23.32 | 48 | 1.31 | 0 | 2 |
| HOPS 60 | 83.84721 | -5.20085 | 420 | 21.93 | 54 | 1.17 | 0 | 2 |
| HOPS 87 | 83.84779 | -5.02464 | 420 | 36.49 | 38 | 1.92 | 0 | 2 |
| HOPS 86 | 83.84854 | -5.02785 | 420 | 3.27 | 113 | 1.49 | I | 2 |
| HOPS 68 | 83.85125 | -5.14183 | 420 | 5.68 | 101 | 0.75 | I | 2 |
| HOPS 78 | 83.85758 | -5.09547 | 420 | 8.93 | 38 | 1.25 | 0 | 2 |
| HOPS 75 | 83.86108 | -5.10286 | 420 | 4.03 | 68 | 0.91 | 0 | 2 |
| HOPS 66 | 83.86183 | -5.15683 | 420 | 20.95 | 265 | 0.07 | flat | 2 |
| HOPS 81 | 83.86646 | -5.08282 | 420 | 1.24 | 40 | 0.85 | 0 | 2 |
| HOPS 85 | 83.86742 | -5.06136 | 420 | 16.28 | 174 | 0.25 | flat | 2 |
| HOPS 96 | 83.87383 | -4.98022 | 420 | 6.19 | 36 | 2.31 | 0 | 2 |
| HOPS 23 | 84.07454 | -5.78180 | 420 | 0.01 | 347 | 0.54 | I | 2 |
| HOPS 176 | 84.09825 | -6.41432 | 420 | 1.52 | 312 | -0.28 | flat | 2 |
| HOPS 185 | 84.15408 | -6.24944 | 420 | 1.04 | 97 | 0.57 | I | 2 |
| HOPS 163 | 84.32200 | -6.60505 | 420 | 0.90 | 432 | 0.36 | I | 2 |
| HOPS 156 | 84.51417 | -6.97106 | 420 | 0.27 | 90 | 1.14 | I | 2 |
| HOPS 135 | 84.68879 | -7.18220 | 420 | 1.14 | 130 | 0.74 | I | 2 |
| HOPS 141 | 84.70008 | -7.01375 | 420 | 0.15 | 742 | -0.06 | flat | 2 |
| HOPS 122 | 84.93804 | -7.32042 | 420 | 0.02 | 246 | 0.80 | I | 2 |
| HOPS 289 | 84.98646 | -7.50169 | 420 | 0.10 | 331 | 0.87 | I | 2 |

| Name | J2000 coordinates, degrees | | $d$, pc | $L_{bol}$, $L_\odot$ | $T_{bol}$, K | $\alpha$ | YSO class | Notes |
|---|---|---|---|---|---|---|---|---|
| HOPS 282 | 85.10871 | -7.62556 | 420 | 0.82 | 95 | 1.86 | I | 2 |
| HOPS 266 | 85.29921 | -7.89327 | 420 | 0.03 | 191 | 0.05 | flat | 2 |
| HOPS 295 | 85.37058 | -2.38871 | 420 | 0.32 | 87 | 1.62 | I | 2 |
| HOPS 303 | 85.51092 | -2.12936 | 420 | 1.49 | 43 | 0.87 | 0 | 2 |
| HOPS 210 | 85.74279 | -8.63483 | 420 | 1.31 | 205 | 2.13 | flat | 2 |
| HOPS 339 | 86.47329 | 0.42425 | 420 | 0.13 | 398 | 0.25 | flat | 2 |
| HOPS 317 | 86.53579 | 0.17736 | 420 | 4.76 | 48 | 0.96 | 0 | 2 |
| HOPS 331 | 86.61800 | 0.33039 | 420 | 0.34 | 83 | 0.25 | flat | 2 |
| HOPS 325 | 86.66354 | 0.02083 | 420 | 6.20 | 49 | 1.19 | 0 | 2 |
| HOPS 322 | 86.69371 | 0.00447 | 420 | 0.48 | 71 | 1.35 | I | 2 |
| HOPS 330 | 86.71404 | 0.32983 | 420 | 0.12 | 385 | 0.28 | flat | 2 |
| HOPS 337 | 86.72958 | 0.39294 | 420 | 0.89 | 129 | 0.81 | I | 2 |
| HOPS 338 | 86.73892 | 0.39729 | 420 | 0.21 | 54 | 1.25 | 0 | 2 |
| HOPS 329 | 86.75671 | 0.29969 | 420 | 2.44 | 89 | 1.31 | I | 2 |
| HOPS 343 | 86.99596 | 0.59246 | 420 | 3.93 | 82 | 1.75 | I | 2 |
| HOPS 1 | 88.55142 | 1.70986 | 420 | 1.52 | 73 | 1.47 | 0 | 2 |
| HOPS 6 | 88.57671 | 1.81762 | 420 | 0.06 | 113 | 1.31 | I | 2 |
| HOPS 7 | 88.58354 | 1.84521 | 420 | 0.53 | 58 | 1.71 | 0 | 2 |
| HOPS 5 | 88.63400 | 1.80199 | 420 | 0.39 | 187 | 0.63 | I | 2 |
| CG 30N | 122.38817 | -36.08279 | 400 | 13.60 | 102 | 0.98 | I | 1, 5 |
| CG 19 | 191.41638 | -55.42278 | 400 | 0.42 | | 1.62 | 0 | 1 |
| SSTc2d J125445.2-770359 | 193.68833 | -77.06651 | 178 | | | 1.16 | I | 1, 3 |
| IRAS 13036-7644 | 196.90363 | -77.00272 | 178 | 0.46 | | > 3 | 0 | 1, 4 |
| SSTc2d J154217.0-524802 | 235.57088 | -52.80063 | 150 | | | 0.82 | I | 1, 3 |
| SSTc2d J154222.5-524806 | 235.59375 | -52.80158 | 150 | | | 0.31 | flat | 1 |
| Lup III MMS | 242.32525 | -39.08140 | 150 | 0.37 | 39 | 1.08 | 0 | 1 |
| SSTc2d J162145.1-234232 | 245.43813 | -23.70884 | 125 | 0.04 | 220 | 0.85 | I | 1 |
| SSTc2d J162159.6-231603 | 245.49821 | -23.26740 | 125 | | | 0.75 | I | 1 |
| SSTc2d J162252.1-243351 | 245.71725 | -24.56424 | 125 | | | -0.02 | flat | 1 |
| SSTc2d J162625.5-242302 | 246.60617 | -24.38376 | 125 | 0.10 | 130 | 0.99 | I | 1 |
| LFAM 4 | 246.60675 | -24.40808 | 125 | 0.04 | 72 | 1.56 | 0 | 1 |
| SSTc2d J162726.3-244246 | 246.85950 | -24.71281 | 125 | 0.02 | 610 | 0.01 | flat | 1 |
| SSTc2d J164404.3-215024 | 251.01792 | -21.84001 | 125 | | | 0.24 | flat | 1 |
| SSTc2d J164526.7-240305 | 251.36113 | -24.05151 | 125 | 0.00 | 500 | 0.44 | I | 1 |
| SSTc2d J182817.6+001607 | 277.07325 | 0.26846 | 260 | | | 0.49 | I | 1 |
| SSTc2d J182846.3+001102 | 277.19308 | 0.18619 | 260 | | | > 3 | 0 | 1 |
| SSTc2d J182853.6+001930 | 277.22350 | 0.32508 | 260 | | | 0.34 | flat | 1 |
| SSTc2d J182854.9+002953 | 277.22867 | 0.49783 | 260 | 2.60 | 51 | 1.56 | 0 | 1 |
| SSTc2d J182854.9+001833 | 277.22883 | 0.30901 | 260 | 0.05 | 120 | 0.64 | I | 1 |
| SSTc2d J182949.6+011706 | 277.45667 | 1.28495 | 260 | 0.33 | 480 | 0.02 | flat | 1 |
| SSTc2d J182951.2+011641 | 277.46325 | 1.27790 | 260 | 1.70 | 140 | 0.56 | I | 1 |

| Name | J2000 coordinates, degrees | | $d$, pc | $L_{bol}$, $L_\odot$ | $T_{bol}$, K | $\alpha$ | YSO class | Notes |
|---|---|---|---|---|---|---|---|---|
| Ser SMM 10 IR | 277.46754 | 1.26322 | 260 | 2.60 | 84 | 1.33 | I | 1 |
| SSTc2d J182952.8+011456 | 277.47021 | 1.24891 | 260 | 1.20 | 120 | 1.03 | I | 1 |
| SSTc2d J182957.6+011301 | 277.49000 | 1.21678 | 260 | 1.40 | 250 | 0.45 | I | 1 |
| SSTc2d J182958.8+011426 | 277.49492 | 1.24061 | 260 | 0.64 | 290 | 0.15 | I | 1 |
| SSTc2d J182959.9+011312 | 277.49975 | 1.21989 | 260 | 2.00 | 130 | 1.68 | I | 1 |

Notes:
(1) E+09. Uncertainty in $L_{bol}$ taken to be 10%.
(2) F+15. Again, the uncertainty in $L_{bol}$ is taken to be 10%.
(3) No mm-submm observations available; the internal luminosity, calculated from the $\lambda = 70$ $\mu$m flux in the manner of Dunham et al. (2008), was used as an estimate of $L_{bol}$. We arbitrarily assign a larger uncertainty, 30%, to these values of $L_{bol}$.
(4) Also Porras et al. 2007, Henning et al. 1993.
(5) Chen et al. 2008.

Table 2: [O I] 63.184 μm observations (Manoj et al. 2013). Upper limits are given as $3\sigma$.

| Name | Line flux, $10^{-13}$ erg cm$^{-2}$ sec$^{-1}$ | Line width (FWHM), km sec$^{-1}$ Observed | Intrinsic* |
|---|---|---|---|
| HOPS 32 | 4.0±0.4 | 134.1 | 87.5 |
| HOPS 10 | 3.2±0.6 | 117.5 | 58.9 |
| HOPS 91 | < 1.9 | | |
| HOPS 56 | 17.2±6.0 | 109.7 | 41.3 |
| HOPS 60 | 39.3±0.8 | 119.2 | 62.2 |
| HOPS 87 | < 1.9 | | |
| HOPS 68 | 2.4±1.0 | 113.9 | 51.3 |
| HOPS 66 | 32.9±1.0 | 114.1 | 51.8 |
| HOPS 85 | 7.2±0.2 | 120.7 | 65.1 |
| HOPS 329 | 1.3±0.4 | 134.3 | 87.7 |
| HOPS 343 | < 0.92 | | |

* After subtracting in quadrature the instrumental width, 101.6 km sec$^{-1}$. We determined this width from high signal/noise observations of objects for which there is independent evidence of a linewidth small compared to the resolution (HOPS 203 = L1641 VLA1; see Manoj et al. 2013). The value we use is in accord with those listed in the *Herschel*-PACS Observer's Manual (http://herschel.esac.esa.int/Docs/PACS/pdf/pacs_om.pdf, section 4.7.1).

Table 3: mid-infrared fine-structure line fluxes from protostars. Upper limits are given as $3\sigma$.

| Name | Line flux ($10^{-15}$ erg sec$^{-1}$ cm$^{-2}$) | | | | |
|---|---|---|---|---|---|
| | [Fe II] 24.5 $\mu$m | [S I] 25.2 $\mu$m | [Fe II] 26.0 $\mu$m | [Si II] 34.8 $\mu$m | [Fe II] 35.3 $\mu$m |
| SSTc2d J032637.5+301528 | 18.8± 0.6 | 7.7± 0.2 | 69.0± 1.2 | 62.8± 2.4 | 14.5± 2.3 |
| L1455 IRS4 | 3.5± 0.3 | 16.3± 1.1 | < 9.7 | 23.1± 1.8 | < 2.6 |
| SSTc2d J032856.6+310737 | 5.1± 0.9 | 19.5± 0.7 | 108.4± 1.2 | 200.6± 3.3 | 32.4± 2.4 |
| SSTc2d J032909.1+312129 | < 7.0 | 7.2± 2.3 | 12.4± 2.6 | 41.1± 7.3 | < 15.4 |
| NGC 1333 IRAS4B | < 1.6 | 198.8± 3.2 | 29.4± 2.4 | 20.4± 0.4 | < 1.5 |
| SSTc2d J032913.6+311358 | 2.0± 0.3 | < 1.4 | 4.7± 0.7 | < 5.9 | < 3.6 |
| SSTc2d J032917.2+312746 | 2.0± 0.2 | 10.5± 1.4 | 12.1± 1.5 | 24.2± 5.9 | < 8.1 |
| SSTc2d J032922.3+311354 | 1.2± 0.2 | < 0.5 | < 0.8 | 3.9± 1.2 | < 2.9 |
| SSTc2d J033327.3+310710 | 6.2± 0.4 | 27.9± 0.6 | 47.7± 1.0 | 64.0± 1.5 | 15.2± 1.0 |
| SSTc2d J034430.3+321135 | < 2.8 | < 1.2 | < 1.8 | < 2.2 | < 7.2 |
| IRAM 04191+15 | < 5.3 | 35.4± 0.4 | 8.7± 1.4 | < 4.7 | < 17.1 |
| HOPS 41 | 9.3± 1.6 | 11.4± 1.5 | < 5.6 | < 7.8 | < 7.8 |
| HOPS 32 | < 2.3 | 11.4± 0.2 | 31.1± 0.4 | 181.0± 2.1 | < 8.9 |
| HOPS 28 | < 2.5 | 2.2± 0.1 | 18.8± 0.5 | 168.5± 2.6 | < 7.8 |
| HOPS 38 | < 2.4 | < 1.3 | 1.9± 0.1 | < 9.9 | < 3.2 |
| HOPS 40 | < 2.1 | < 2.5 | < 2.6 | < 8.8 | < 5.9 |
| HOPS 10 | 14.1± 1.0 | 18.0± 0.9 | 127.2± 0.1 | 160.5± 6.0 | 33.6± 5.4 |
| HOPS 371 | < 0.3 | < 0.3 | 2.7± 0.7 | < 3.4 | < 4.6 |
| HOPS 91 | < 3.3 | < 2.9 | 1.7± 0.5 | < 26.2 | < 14.1 |
| HOPS 56 | < 22.0 | 51.6± 7.5 | 32.9± 3.8 | 35.2± 6.9 | < 24.1 |
| HOPS 60 | 8.8± 2.4 | 93.8± 13.3 | 217.2± 13.6 | 255.7± 82.1 | < 228.0 |
| HOPS 87 | 9.4± 2.5 | 159.1± 1.0 | < 9.7 | < 17.2 | < 28.7 |
| HOPS 86 | 3.2± 1.0 | 31.5± 1.6 | 15.2± 0.9 | 26.3± 8.3 | < 12.3 |
| HOPS 68 | < 4.1 | 47.3± 6.3 | 22.1± 3.9 | 57.8± 11.7 | < 7.4 |
| HOPS 78 | 33.9± 5.8 | 38.2± 6.2 | 157.9± 11.9 | 251.6± 30.7 | 63.4± 10.4 |
| HOPS 75 | 19.6± 4.9 | < 18.6 | < 18.4 | < 30.3 | < 21.4 |
| HOPS 66 | 72.2± 5.3 | 110.9± 5.5 | 322.9± 7.4 | 488.5± 14.6 | 82.5± 15.9 |
| HOPS 81 | 10.6± 0.7 | 20.2± 1.4 | < 0.4 | < 14.8 | < 17.1 |
| HOPS 85 | < 11.6 | 38.4± 5.2 | 32.4± 5.0 | 71.4± 18.4 | < 35.0 |
| HOPS 96 | < 4.1 | < 6.4 | < 5.7 | < 11.1 | < 9.0 |
| HOPS 23 | < 1.3 | < 0.9 | < 1.5 | < 10.4 | < 7.0 |
| HOPS 176 | < 3.3 | < 3.9 | 8.0± 0.7 | 30.4± 4.8 | < 5.5 |
| HOPS 185 | < 2.5 | 2.1± 0.1 | 1.1± 0.1 | 7.7± 0.5 | < 1.2 |
| HOPS 163 | < 1.9 | 2.3± 0.6 | < 1.2 | < 9.3 | < 7.2 |
| HOPS 156 | < 1.1 | 2.7± 0.2 | < 0.6 | < 3.7 | 5.4± 1.6 |
| HOPS 135 | 4.8± 0.8 | 3.9± 0.2 | 18.0± 0.3 | 21.3± 3.3 | 4.5± 0.9 |
| HOPS 141 | < 2.4 | < 3.4 | < 4.2 | < 6.4 | < 4.1 |
| HOPS 122 | < 1.2 | < 0.9 | < 0.7 | < 3.4 | < 2.2 |

| Name | Line flux ($10^{-15}$ erg sec$^{-1}$ cm$^{-2}$) | | | | |
|---|---|---|---|---|---|
| | [Fe II] 24.5 $\mu$m | [S I] 25.2 $\mu$m | [Fe II] 26.0 $\mu$m | [Si II] 34.8 $\mu$m | [Fe II] 35.3 $\mu$m |
| HOPS 289 | < 3.7 | < 2.2 | 5.3± 0.6 | < 30.1 | < 32.3 |
| HOPS 282 | < 2.6 | 3.6± 0.4 | 23.4± 1.0 | 26.2± 0.3 | 7.3± 1.6 |
| HOPS 266 | 2.0± 0.1 | < 0.8 | < 0.6 | < 4.2 | < 3.1 |
| HOPS 295 | < 0.7 | < 1.2 | 3.2± 0.4 | < 3.5 | < 2.1 |
| HOPS 303 | 2.9± 0.1 | 1.9± 0.2 | 32.5± 0.5 | 41.3± 0.6 | 12.9± 1.9 |
| HOPS 210 | < 3.1 | < 5.1 | < 3.9 | 3.4± 1.1 | < 3.7 |
| HOPS 339 | < 0.7 | < 0.6 | < 0.3 | 2.6± 0.5 | < 2.6 |
| HOPS 317 | < 3.1 | 14.6± 1.3 | 54.9± 1.4 | 80.4± 5.5 | < 15.1 |
| HOPS 331 | < 1.2 | < 0.6 | < 1.2 | < 6.4 | < 2.7 |
| HOPS 325 | 27.7± 1.8 | 73.5±13.9 | 422.5± 2.2 | 258.3±22.2 | 91.9±24.3 |
| HOPS 322 | < 21.5 | < 24.4 | < 25.9 | < 57.0 | < 57.8 |
| HOPS 322 | < 1.1 | 2.0± 0.1 | 3.5± 1.1 | 11.1± 3.5 | < 4.9 |
| HOPS 330 | < 0.3 | < 1.5 | 1.4± 0.2 | < 1.0 | < 1.5 |
| HOPS 337 | < 2.0 | 3.5± 0.5 | 10.8± 0.2 | 16.5± 0.3 | < 4.7 |
| HOPS 338 | < 0.1 | 4.1± 0.3 | 5.8± 1.4 | 10.9± 2.7 | < 3.9 |
| HOPS 329 | < 7.9 | 7.4± 2.1 | 11.6± 0.1 | 50.9± 5.0 | < 6.4 |
| HOPS 343 | < 4.7 | 20.0± 1.9 | 10.2± 1.7 | < 16.1 | < 11.8 |
| HOPS 1 | 3.5± 0.2 | 4.4± 0.2 | 43.4± 0.3 | 29.7± 0.4 | 6.9± 2.1 |
| HOPS 6 | 0.7± 0.2 | 1.0± 0.2 | < 0.4 | < 2.9 | 4.6± 0.7 |
| HOPS 7 | < 0.8 | 1.7± 0.2 | < 1.6 | < 7.0 | < 4.1 |
| HOPS 5 | < 0.9 | < 0.2 | 0.8± 0.2 | < 2.3 | < 3.5 |
| CG 30N | < 10.3 | 40.9± 4.6 | 37.4± 4.4 | 73.8±15.2 | < 36.1 |
| CG 19 | < 1.3 | 5.7± 0.2 | 6.5± 0.9 | 18.3± 2.5 | < 2.7 |
| SSTc2d J125445.2-770359 | < 0.8 | < 0.8 | 0.9± 0.2 | < 4.5 | < 4.9 |
| IRAS 13036-7644 | 4.2± 1.0 | 22.8± 1.2 | 29.7± 0.8 | 80.7± 2.6 | < 7.0 |
| SSTc2d J154217.0-524802 | < 1.4 | < 1.9 | 5.0± 0.7 | < 11.5 | < 3.9 |
| SSTc2d J154222.5-524806 | < 0.6 | < 0.7 | < 0.2 | < 8.0 | < 2.5 |
| Lup III MMS | 7.3± 0.4 | 2.5± 0.1 | 1.5± 0.1 | 5.2± 1.4 | < 18.5 |
| SSTc2d J162145.1-234232 | 3.4± 0.1 | 7.0± 0.2 | 10.2± 0.3 | 11.2± 0.9 | < 4.1 |
| SSTc2d J162159.6-231603 | < 0.9 | 1.0± 0.2 | < 0.8 | < 4.7 | < 7.6 |
| SSTc2d J162252.1-243351 | < 0.7 | < 0.2 | < 1.6 | < 7.2 | < 8.2 |
| SSTc2d J162625.5-242302 | < 2.4 | 4.0± 0.3 | < 7.0 | 10.0± 0.9 | < 8.3 |
| LFAM 4 | < 2.7 | 7.0± 0.1 | 18.6± 0.2 | 45.1± 4.0 | < 10.2 |
| SSTc2d J162726.3-244246 | 1.1± 0.2 | < 0.8 | < 0.7 | < 1.6 | < 3.8 |
| SSTc2d J164404.3-215024 | < 0.7 | < 0.6 | < 1.1 | < 2.0 | < 4.2 |
| SSTc2d J164526.7-240305 | < 0.6 | < 0.6 | < 0.7 | < 3.3 | < 2.6 |
| SSTc2d J182817.6+001607 | < 0.6 | 1.1± 0.3 | < 0.7 | < 2.8 | < 2.3 |
| SSTc2d J182846.3+001102 | < 1.0 | < 0.5 | < 0.7 | < 1.7 | < 3.1 |
| SSTc2d J182853.6+001930 | 0.8± 0.2 | < 0.7 | 0.3± 0.1 | 7.5± 1.1 | < 3.3 |
| SSTc2d J182854.9+002953 | 4.4± 0.3 | 16.2± 0.7 | 22.4± 0.2 | 48.0± 5.3 | < 7.7 |
| SSTc2d J182854.9+001833 | < 1.5 | < 1.1 | < 1.2 | < 2.6 | < 3.8 |

| Name | Line flux ($10^{-15}$ erg sec$^{-1}$ cm$^{-2}$) | | | | |
|---|---|---|---|---|---|
| | [Fe II] 24.5 $\mu$m | [S I] 25.2 $\mu$m | [Fe II] 26.0 $\mu$m | [Si II] 34.8 $\mu$m | [Fe II] 35.3 $\mu$m |
| SSTc2d J182949.6+011706 | 6.1 ± 0.4 | 29.8 ± 5.7 | 22.9 ± 0.9 | 62.6 ± 11.1 | < 8.9 |
| SSTc2d J182951.2+011641 | 15.8 ± 0.2 | 2.3 ± 0.6 | 107.0 ± 0.4 | 194.2 ± 1.2 | 44.6 ± 1.8 |
| Ser SMM 10 IR | 8.7 ± 1.3 | 30.4 ± 0.2 | < 3.8 | < 14.8 | < 15.6 |
| SSTc2d J182952.8+011456 | < 5.7 | 30.4 ± 0.4 | < 17.5 | < 13.4 | < 14.2 |
| SSTc2d J182957.6+011301 | 12.0 ± 1.8 | 15.9 ± 0.7 | 14.9 ± 2.4 | 44.7 ± 4.4 | < 24.3 |
| SSTc2d J182958.8+011426 | < 6.2 | 19.3 ± 1.6 | 16.3 ± 1.7 | < 30.2 | < 18.1 |
| SSTc2d J182959.9+011312 | 18.5 ± 1.3 | 33.2 ± 1.4 | < 15.7 | < 11.8 | < 6.4 |

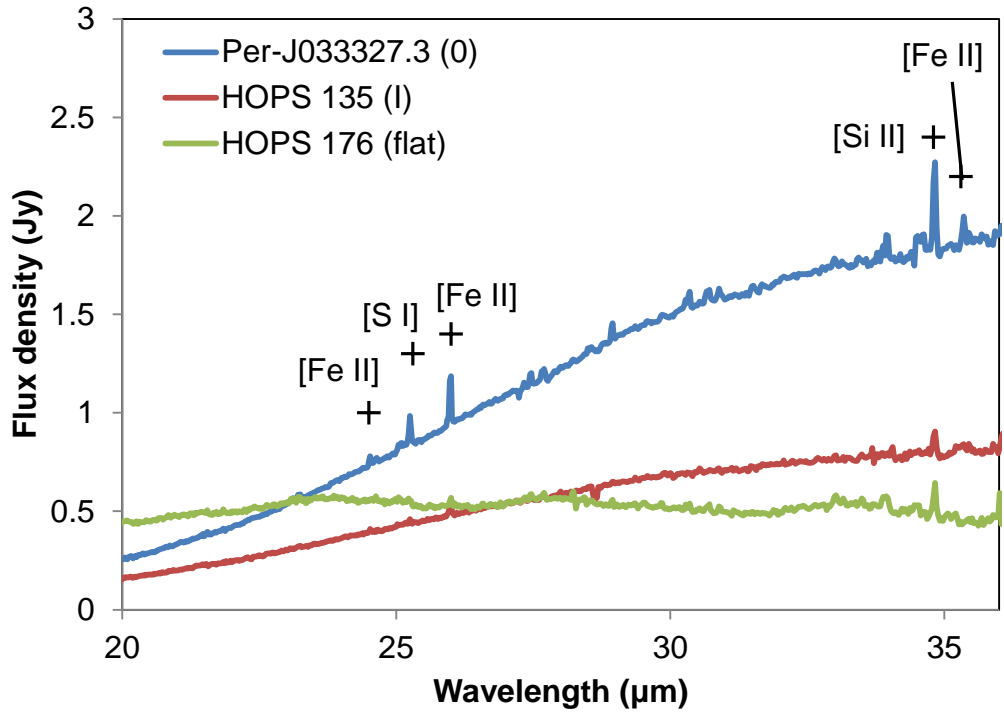

Figure 1: *Spitzer*-IRS LH spectra of typical Class 0, I and flat-spectrum YSOs in our sample. Wavelengths of emission lines of [Si II], [Fe II] and [S I] are labelled. Numerous pure-rotational lines of $H_2O$ and OH are also detected in each spectrum; these will be discussed in a subsequent article.

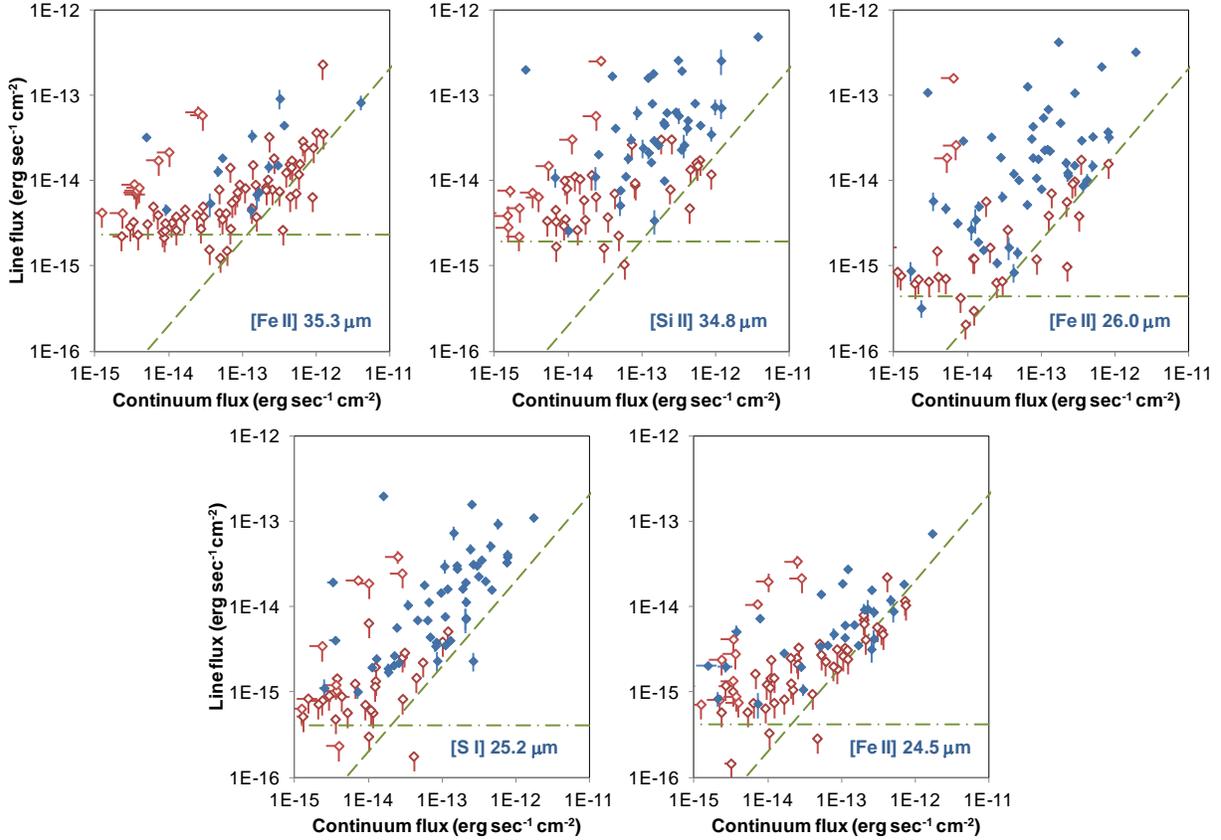

Figure 2: Fine-structure line detections and upper limits in our protostar sample, compared to the survey detection limits. Each frame is a plot of line flux against the flux per spectral resolution element in the continuum beneath the line, for the spectral line indicated in the lower right of the frame. Solid (blue) symbols indicate that both line and continuum were detected. Open (red) symbols indicate that one or both of the line and continuum were not detected, with 3σ values plotted for upper limits. Error bars are ±1σ for detections, -1σ for upper limits. The dot-dash (green) horizontal line gives the formal 3σ survey detection limit based on *Spitzer*-IRS sensitivity. The long-dash (green) line indicates the IRS flat-field limit according to instrument specifications.

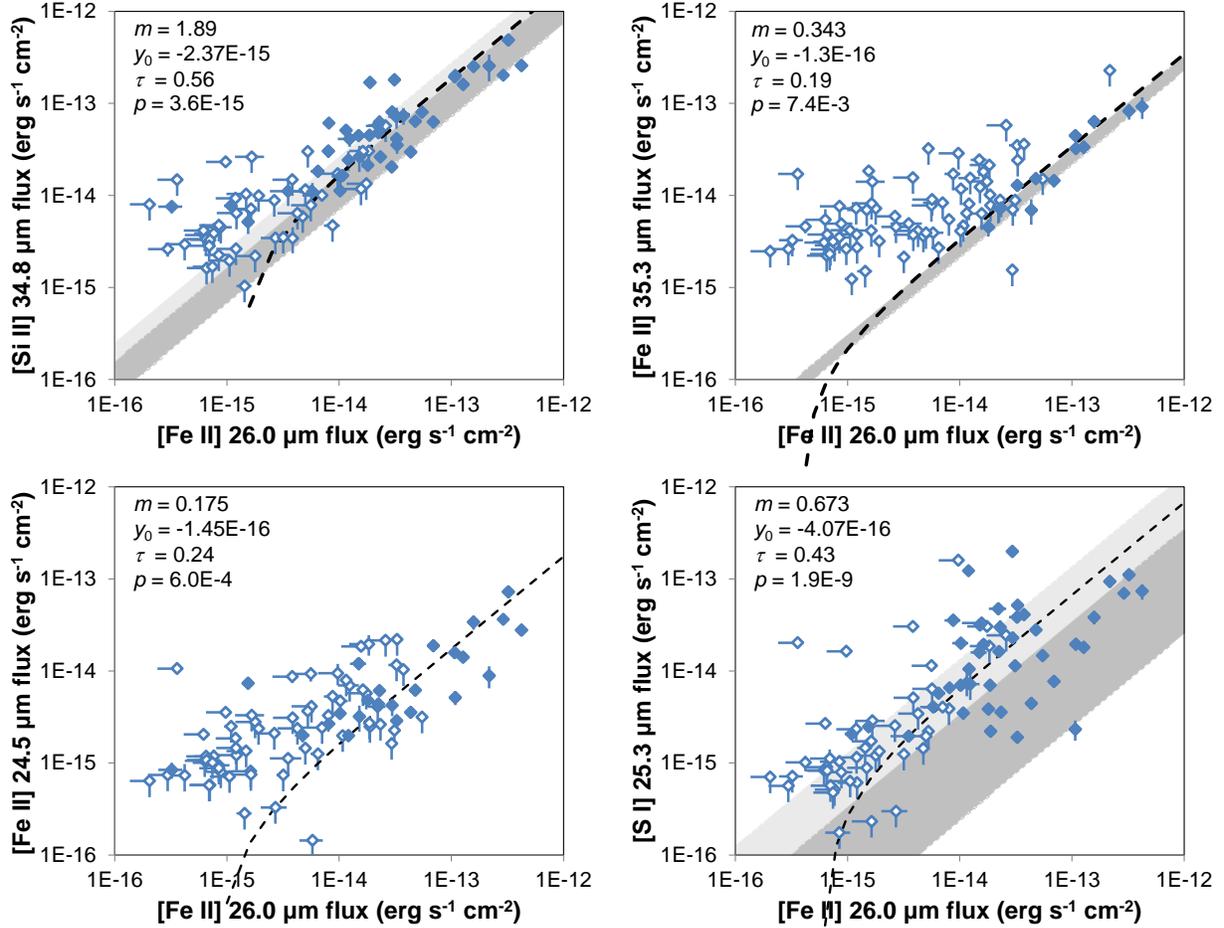

Figure 3: Comparison of measured fine-structure line flux ratios to models for line emission by gas behind J shocks. Solid (blue) symbols indicate detection of both lines in the pair plotted. Open (blue) symbols indicate non-detection of one or both lines, with $3\sigma$ values plotted for the upper limits. Error bars are $\pm 1\sigma$ for detections, $-1\sigma$ for upper limits. The dashed lines are linear regressions to the data – despite the log-log plots – using the Akitras-Thiel-Sen (henceforth ATS) survival-analysis method (e.g. Feigelson & Babu 2012). Regression statistics are given in the upper left of each frame: slope $m$, intercept $y_0$, Kendall correlation coefficient $\tau$, and the probability $p$ that a correlation of the value $\tau$ would occur at random in a sample of the same size. Gray shading indicates the domain occupied by the J shock models by HM89 for preshock hydrogen-atom density $n_0 = 10^3 - 10^4$ cm$^{-3}$ and shock speed $v_S < 40$ km sec$^{-1}$ (light) and $v_S = 40 - 150$ km sec$^{-1}$ (dark). The [Fe II] 24.5 µm line was not included in the HM89 calculations.

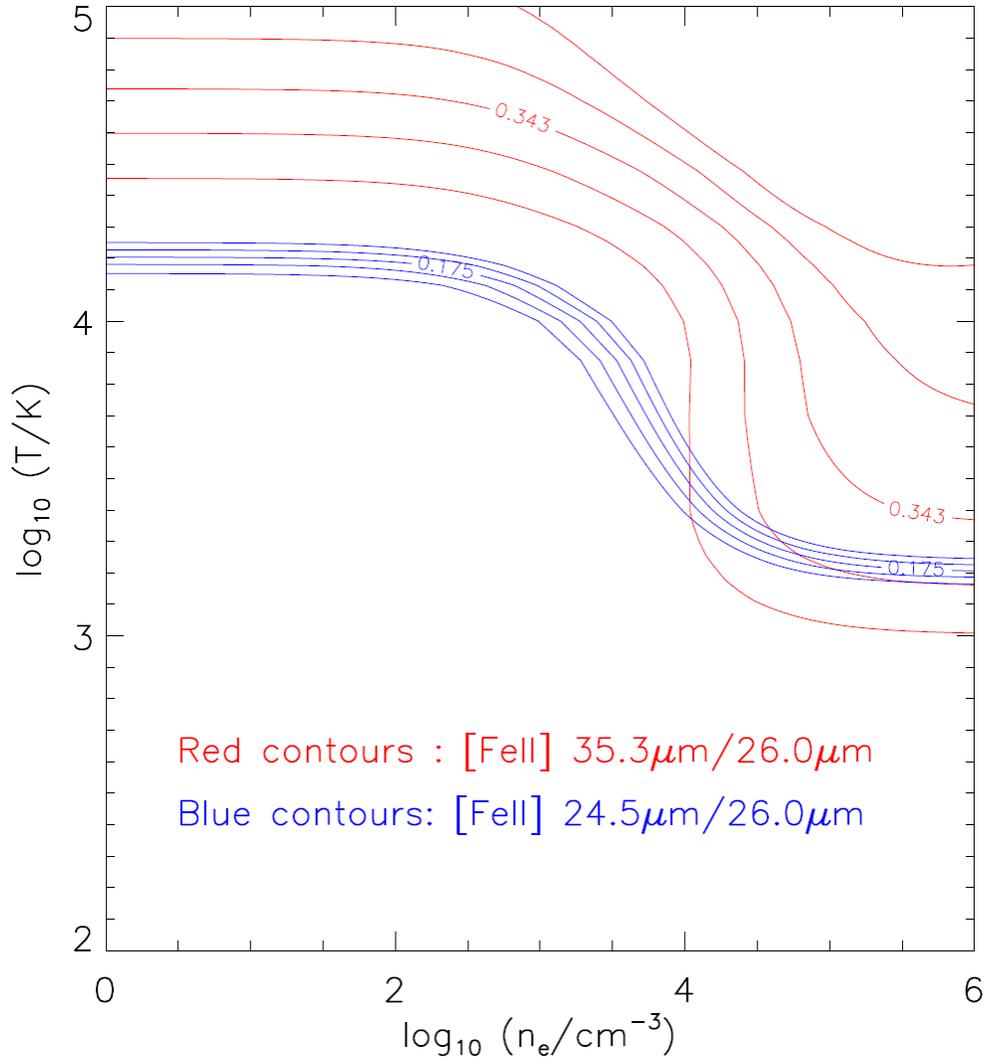

Figure 4: Mid-infrared [Fe II] line intensity ratios in a single-component model, as functions of density and temperature. The lowest 63 energy levels of Fe$^+$ were considered in the calculation, and the collisional excitation rate coefficients calculated by Ramsbottom et al. (2007) were used, along with A-coefficients from Nussbauer & Storey (1988). The upper (red) and lower (blue) sets of contours are respectively the intensity ratios 35.3/26.0 $\mu$m and 24.5/26.0 $\mu$m. Contour levels are set at the values given by the regressions in Figure 3, and at ±10% and ±20% of these values.

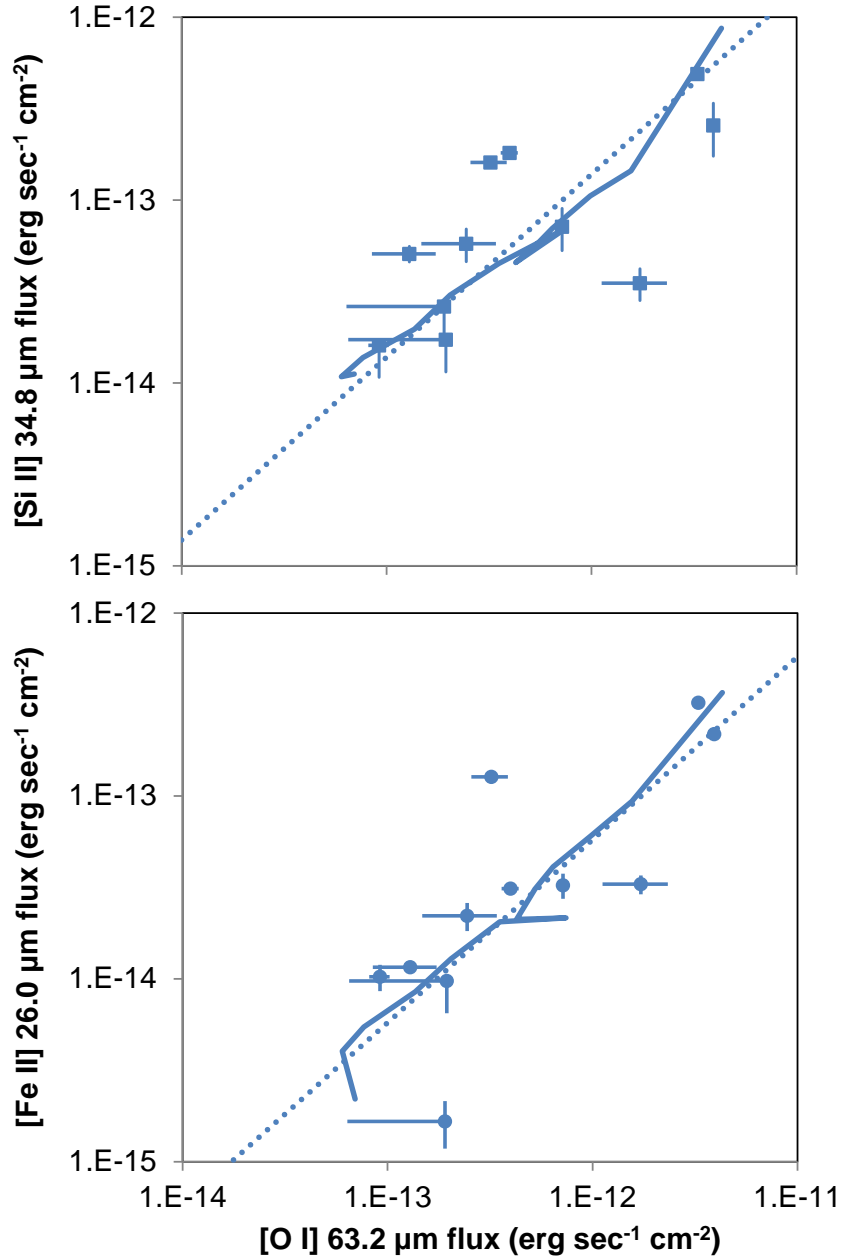

Figure 5: Observed fluxes of [Si II] 34.8 $\mu$m, [Fe II] 26.0 $\mu$m, and [O I] 63.2 $\mu$m lines (solid symbols), compared to predictions by the *J* shock models by HM89 (solid lines) for preshock nucleon density range $n_0 = 10^3 - 10^4$ cm$^{-3}$, shock speed range $v_w$ = 30-150 km sec$^{-1}$, and an arbitrarily-chosen beam filling factor of 7%. (The models move along 45° in the plot, if the filling factor is adjusted.) Dotted lines are ATS regressions to the observations. The models are plotted for $Si^+/O = 6.7 \times 10^{-3}$ and $Fe^+/O = 1.9 \times 10^{-3}$, the latter as in HM89, the former scaled by a least-squares fit to the observations.

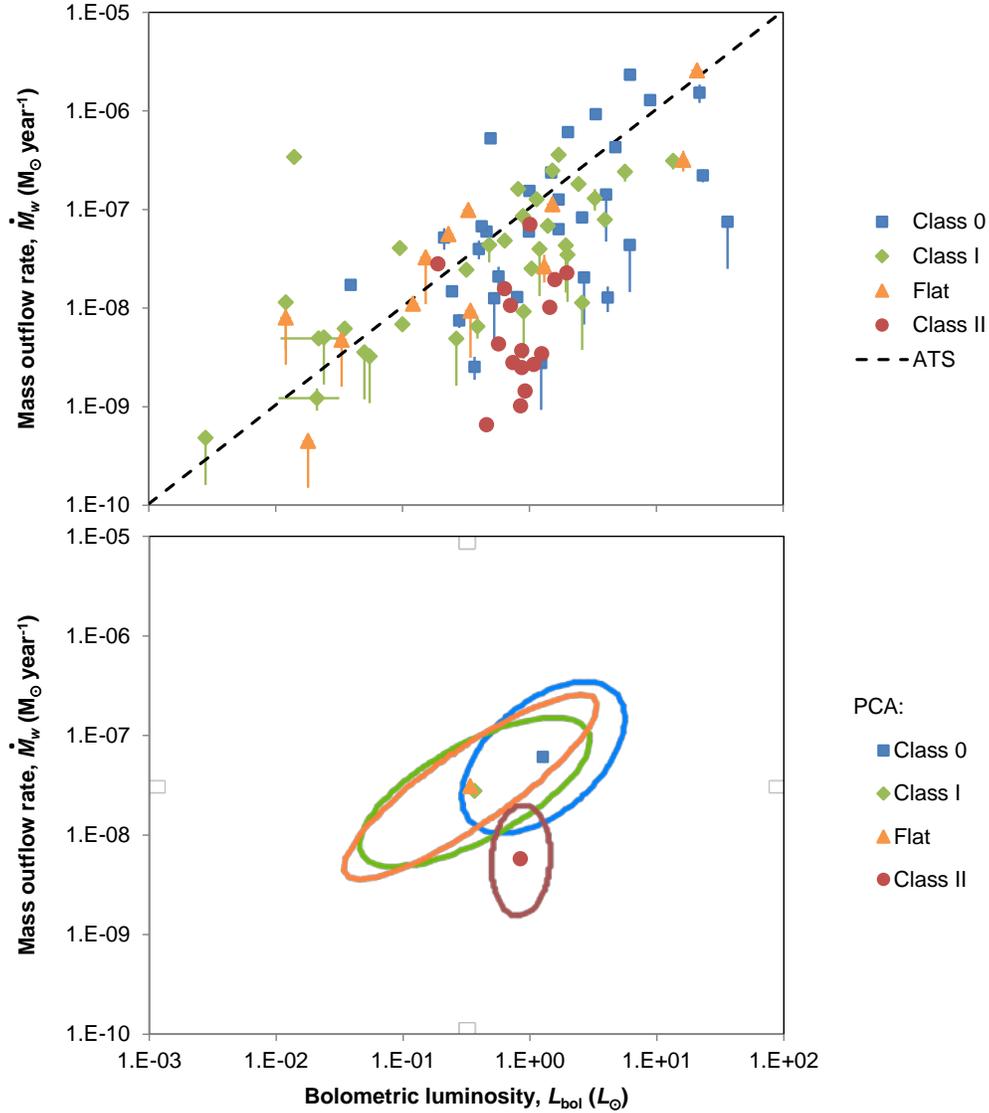

Figure 6: Mass outflow rate $\dot{M}_w$ vs. bolometric luminosity $L_{bol}$, for our sample of protostars and for Class II objects characterized by HEG95 and G+98. *Top:* Class 0 objects plotted as blue squares, Class I as green diamonds, flat-spectrum as orange triangles, Class II as red circles. Upper limits are plotted as $3\sigma$ points with $2\sigma$ tails; all other points are plotted with $\pm 1\sigma$ errorbars. The dashed line is an ATS linear regression through the origin to the Class 0/I/flat-spectrum data, with $m = 1.04 \times 10^{-7} M_\odot$ year$^{-1}$ $L_\odot^{-1}$, $\tau = 0.42$, and $p = 6.1 \times 10^{-8}$. *Bottom:* principal components analysis, with mean values plotted in the same symbols as above, and first and second principal-component eigenvectors indicated respectively by major and minor axes of ellipses centered on the corresponding mean value.

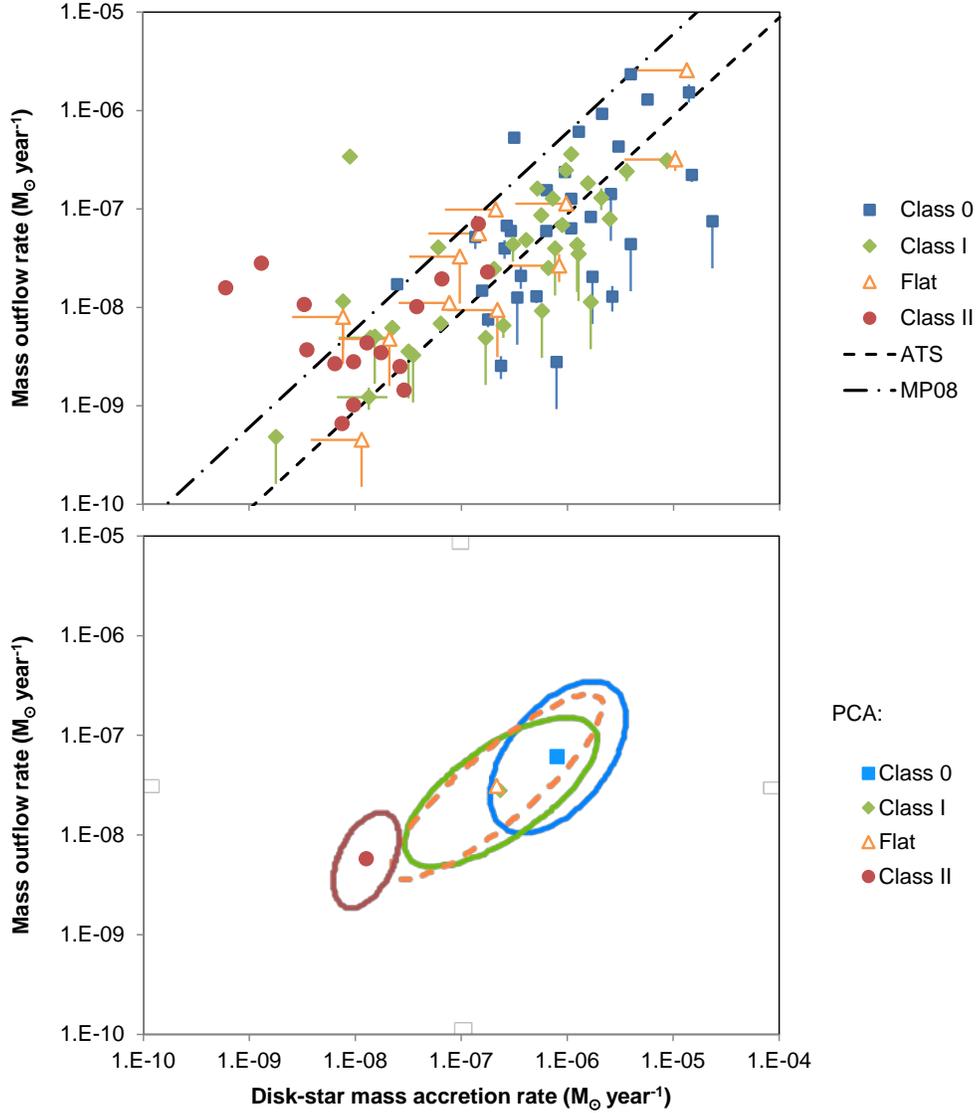

Figure 7: Same as Figure 6, but for mass outflow rate $\dot{M}_w$ vs. disk-star accretion rate. For flat spectrum objects, the upper limits to $\dot{M}_a$ resulting from Equation (3) are plotted as empty triangles with factor-of-1.5 tails. The dashed line is an ATS linear regression through the origin to the Class 0/I/flat-spectrum data, with $m = 0.089$, and other statistics the same as in Figure 6. The black dot-dash line near the upper edge of the distribution of data is $\dot{M}_w = 0.6\dot{M}_a$, which corresponds to the maximum branching ratio possible for an accretion-powered stellar wind (MP08).

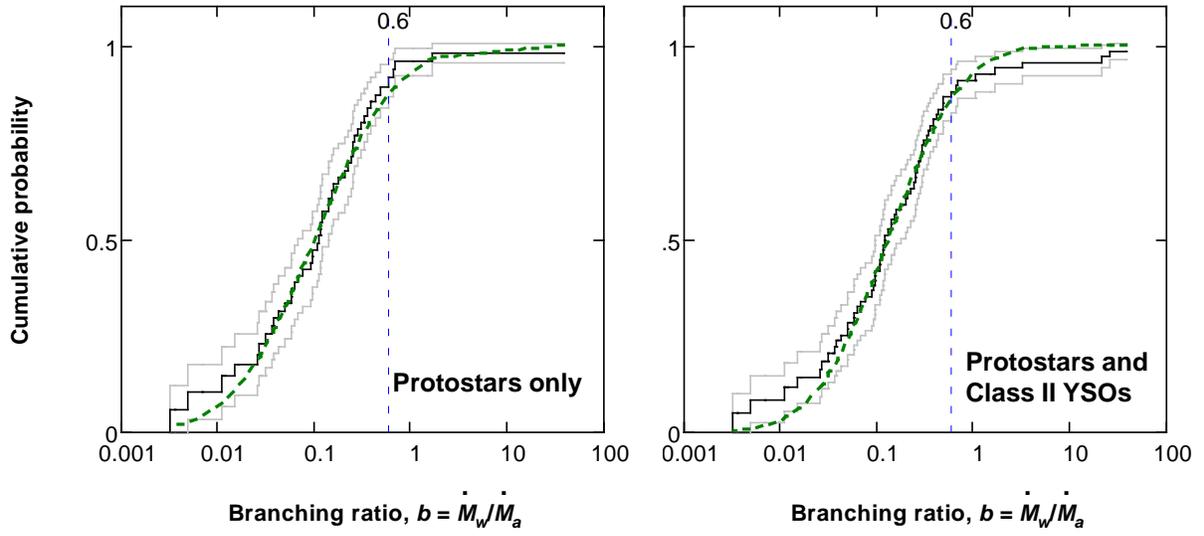

Figure 8: Kaplan-Meier-Efron estimate of the empirical distribution function for $b = \dot{M}_w / \dot{M}_a$ (black steps), and the 95% confidence limits for this estimate (gray steps), for the protostars in our survey (left) and for both protostars and Class II objects (right). Overlaid is the cumulative distribution function for a lognormal probability density (green dashes), fit to the branching-ratio EDF in the range $b$ = 0.05-0.5 by $\chi^2$ minimization. A marker at $b$ = 0.6 (blue dots) indicates the upper limit to $b$ in APSW outflow models, which also corresponds approximately to the upper edge to the $b$ distribution.